# TOWARDS AUTONOMIC ORCHESTRATION OF MACHINE LEARNING PIPELINES IN FUTURE NETWORKS

Abhishek Dandekar[1]
[1]TU Berlin, Germany,

NOTE: Corresponding author: Abhishek Dandekar, dandekar@campus.tu-berlin.de

**Abstract** – *Machine learning (ML) techniques are being increasingly used in mobile networks for network planning, operation, management, optimisation and much more. These techniques are realised using a set of logical nodes known as ML pipeline. A single network operator might have thousands of such ML pipelines distributed across its network. These pipelines need to be managed and orchestrated across network domains. Thus it is essential to have autonomic multi-domain orchestration of ML pipelines in mobile networks. International Telecommunications Union (ITU) has provided an architectural framework for management and orchestration of ML pipelines in future networks. We extend this framework to enable autonomic orchestration of ML pipelines across multiple network domains. We present our system architecture and describe its application using a smart factory use case. Our work allows autonomic orchestration of multi-domain ML pipelines in a standardised, technology agnostic, privacy preserving fashion.*

## 1. INTRODUCTION

In the last decade Machine Learning (ML) and Artificial intelligence (AI) have seen tremendous growth ranging from facial recognition and smart personal assistants to cancer detection. With increasingly capable hardware, availability of big data and novel algorithmic techniques AI/ML is being seen as a solution to solve complex problems in various domains like autonomous cars, healthcare, disaster management and climate change. One of these domains is mobile networks and especially 5th generation mobile networks (5G).

5G has introduced new features which have enabled new use cases across various verticals. In the previous generation of mobile networks majority of end users/end points were human. However, in the newer 5G and beyond use cases, majority of end points are machines. It is estimated that number of connected machines will be three times more than number of connected humans[1]. This has created new challenges and opportunities. In order to handle these challenges, a more complex, robust and scalable mobile network is required. At the same time, huge amounts of data is now available for analytics from all the connected machines. This has created a unique opportunity to apply AI/ML techniques to 5G networks.

In order to make most of this opportunity the International Telecommunications Union (ITU) launched ITU AI/ML in 5G Challenge 2020. This provided a platform for the community to innovate and solve problems related to 5G networks using AI/ML. We participated in this challenge and proposed a solution based on the ITU standard Y.3172[2]. This standard defines a framework to orches-trate ML pipelines in 5G networks. We implemented a minimal version of Machine Learning Function Orchestrator (MLFO) which demonstrated intent parsing, model selection, interaction with ML marketplace and ML underlay[3].

Other participants in this challenge proposed ML based solutions to various problems like traffic recognition, link failure prediction, channel estimation etc.[4]. We found two notable issues across all the presented solutions. Firstly, a common problem which was echoed by the participants and keynote speakers alike was that the telecom companies and industries in other verticals were hesitant to share the network related data and did not want to reveal their internal complexities. Secondly, most of the solutions needed human expertise to decide *when* and *where* would the solutions be deployed in the network. They also needed human intervention in case they needed to be updated, optimised or redeployed.

Integration of any ML based solutions in mobile networks requires a mechanism which can deploy, update and tear down ML pipelines. This mechanism is called ML pipeline orchestration. In order to deploy solutions proposed in the AI/ML in 5G challenge – firstly, it is necessary for the ML pipeline orchestration to be autonomic, this means that ML pipelines should be able to orchestrate themselves. Secondly, multi-domain ML pipeline orchestration is required as the ML based solution might also involve multiple pipelines across different network or operator domains. This requires use of standardised interfaces across domains. Thirdly, owners of these domains might not be willing to share private data and expose in-ternal technological complexities of the domain. Hence the ML pipeline orchestration needs to be privacy preserving and technology agnostic.

We propose a framework which allows autonomic orchestration of ML pipelines. We achieve this by extending the framework for orchestration of ML pipelines in future networks defined in ITU Y.3172[2]. Rest of the paper is organised as follows. Section 2 describes relevant research work and standardisation activities. Section 3 provides relevant background and describes the system design. Section 4 provides application of our framework in smart factory use case. Section 5 provides conclusion.

## 2. RELATED WORK

In this section we review the research literature and then discuss relevant work done in various Standard Development Organisations (SDOs).

[5] presents the fundamental concepts of autonomic computing and describes their major features like self-configuration, self-healing, self-optimisation, self-protection. [6] provides classification of various autonomic network architectures like hierarchical networks, flat networks and provides evaluation criteria for each of the architectures. [7] provides classification of various autonomic network management architectures according to their degree of autonomicity. [8][9] provide an overview of various state of the art AutoML frameworks used for automatic construction of ML pipelines.

European Telecommunications Standards Institute (ETSI) has proposed Experiential Networked Intelligence (ENI) framework[10]. ENI enables network to change its behaviour with changing business, environmental and end user conditions. ENI achieves this by using AI/ML techniques on data gathered from the network subsystems. ENI reduces need of human intervention while also assisting humans in decision making process. ETSI has also defined Generic Autonomic Networking Architecture (GANA) which is a reference model for autonomic and cognitive networking[11]. It realises autonomic features like self-configuration, self-optimisation by tuning the configurable parameters in the network elements or in the higher layer Operation and maintenance (OAM) functions.

TM Forum has proposed their version of autonomous network which aims to provide a zero-touch, zero-wait network service[12]. The network is divided into various autonomous domains and each autonomous domain can perform closed loop automation according to the requirements. Multi-domain closed loop automation can also be achieved by using an intent.

ITU has provided several frameworks to enable autonomic networks. ITU Y.3172[2] defines a framework for machine learning in International Mobile

Telecommunications-2020 (IMT-2020) networks. ITU Y.3173[13] describes five levels of network intelligence and a methodology to evaluate these levels. ITU Y.3324[14] provides high level functional requirements for autonomic management and control of IMT-2020 networks. It also proposes a high level architecture to provide autonomic management and control capability for control, data and management planes. ITU study group SG13 has recently established a focus group on autonomous networks for pre-standard study of enablers of autonomous networks[15].

Internet Engineering Task Force (IETF) in [16]and [17]provides overview of intent based networking and how it is used in context of autonomic networks.

## 3. SYSTEM DESIGN

### 3.1 Background

In this subsection we briefly describe concepts which are fundamental to understand our system architecture.

#### 3.1.1 ITU framework for ML in future networks

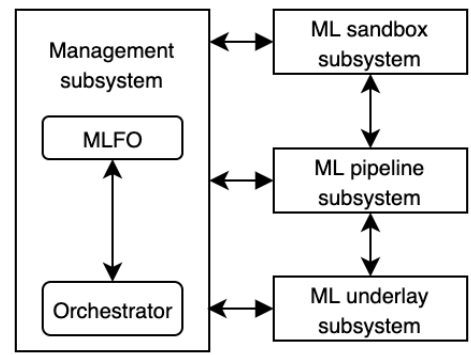

Fig. 1 – ITU framework for ML in future networks

Figure 1 (derived from fig.3 in[2] ) shows the high level architecture of the ITU framework. It consists of four major architectural components-

1. Management subsystem: This subsystem is responsible for management and orchestration of ML and Non-ML functions in the IMT-2020 networks. The Non-ML functions are orchestrated by the traditional Operations, administration and management (OAM) orchestrators. The ML function orchestration is carried out by Machine Learning Function Orchestrator (MLFO) in conjunction with the OAM orchestrator. MLFO has four major functionalities, (1) Intent parsing, (2) Lifecycle management of ML pipelines, (3) Management of data and configuration of data sources, (4) Management of ML models which may include model selection, training, deployment etc. Mobile network operator may specify which ML application it wants to deploy using a input *intent* to the MLFO.

2. ML Pipeline subsystem: An ML application can logically be viewed as a chain of logical nodes i.e. a pipeline. This pipeline consists of data source, collector, preprocessor, model, policy and data sink. When the MLFO receives an intent it deploys a corresponding ML pipeline in the network infrastructure.

3. ML Underlay: ML underlay is the IMT-2020 telecommunication network on which the ML applications are to be deployed. The management subsystem interacts with the underlay to deploy the ML pipeline in the network. Examples of ML underlay are 5G, 6G, IEEE 802.11 etc.

4. ML sandbox subsystem: This is used to test and evaluate ML pipelines before deploying them on live network. This allows ML pipelines to adapt to dynamic network environments.

### 3.1.2   Autonomic computing

Computing systems which can manage themselves (self-management) based on high level objectives from administrators are termed autonomic computing systems[5]. Self-management is defined by the following four capabilities, also known as self-* or self-CHOP capabilities.

- Self-configuration: The capability of the system to configure its own components by following the high-level policy or intent.

- Self-healing: The capability of the system to detect, diagnose and fix problems in its components.

- Self-optimisation: The capability of the system to continuously monitor its components to optimise its performance and efficiency.

- Self-protection: The capability of the system to defend itself from malicious attacks and prevent large scale failures.

In scope of this paper we limit ourselves to self-configuration.

## 3.2   Hierarchical MLFO

A communication network can be divided into several network sub-systems. Each sub-system can be treated as an network domain. Network domains can be logical domains like Radio Access Network (RAN), Core Network (CN), Transport Network (TN) or administrative domains like network operated by mobile operator, network operated by a industry vertical, network operated by mobile virtual network operator etc. Each such domain might host one or more ML pipelines. In such multi-domain scenarios, it is essential to have the ability to orchestrate pipelines across domains.

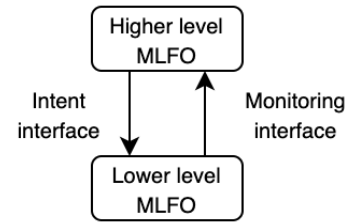

Fig. 2 – Hierarchical MLFO

Network domains typically have a control hierarchy where domains higher in hierarchy set policy or configurations for domains lower in hierarchy. Figure 2 shows that MLFOs can be arranged in a similar control hierarchy by assigning one MLFO per network domain. These MLFOs communicate using two interfaces, (1) an intent interface which is used by higher level MLFOs to apply a set of requirements to lower level MLFOs, (2) a monitoring interface which is used to get performance analytics data from lower level MLFOs to higher level MLFOs. Intent can be of two types- (1) operator intent and (2) MLFO intent. Operator intent consist of high level requirements from the mobile network operator and is applied to higher level MLFO. MLFO intent is derived by higher level MLFO based on analytics provided by ML pipelines within its domain and the requirements provided by the operator. This intent is then sent over the intent interface to lower level MLFO. The lower level MLFO then parses and interprets the intent and orchestrates the corresponding ML pipelines. The lower level MLFO sends asynchronous updates to higher MLFOs about the performance of the ML pipelines. In this way, the hierarchical structure of MLFOs allows ML pipelines to be orchestrated without any human intervention. They can also be optimised and reconfigured on the fly in response to new set of requirements or changes in network conditions. This makes the ML pipeline orchestration mechanism capable of self-configuration.

In the following subsections we describe the intent and monitoring interface in detail.

## 3.3   Intent interface

Intent is an declarative, technology agnostic description of configuration and policy requirements for ML pipeline. Instead of exactly instructing the MLFO *how* ML pipeline should perform its function, intent only instructs *what* outcome is expected from the pipeline along with some constraints. Using intent based mechanism has couple of advantages. Firstly, domains do not need to reveal their internal complexities or proprietary ML mechanisms to other domains as intent is technology agnostic. Secondly, it provides more autonomy to lower domains to take low level decisions. The intent consists of following fields-

- *intentid* is a unique identifier for every intent

- *targetid* specifies identity of the network domain (e.g. Fog-001, Edge-001)

- *operation* specifies *what* action is to be performed (e.g maximise, minimise)

- *operand* specifies to *which* parameter the action is to be applied (e.g. Quality of service (QoS), accuracy, capacity)

- *oparams* specifies operational parameters necessary to perform the action (e.g. 5QI=1 specifies required QoS level[18])

- *constraints* specifies which constraints need to be followed while performing the action (e.g. latency limit, privacy level)

We provide examples of intents in section 4.2

## 3.4 Monitoring interface

This interface is used for collecting performance data of ML pipelines from lower level MLFO. The performance data might include statistics about model accuracy, resource utilisation, available resources etc. Monitoring interface may also be used to send event based asynchronous updates. These events may include threshold crossover, policy violation, failure event, security violation event etc.[19]. Based on the analysis of data collected from the monitoring interface, the higher level MLFO may create a new set of requirements and send a MLFO intent over intent interface to reconfigure or optimise the ML pipeline.

## 4. USE CASE: SMART FACTORY

In this section we present a practical application of our hierarchical MLFO architecture. We consider a smart factory use case and demonstrate how self-configuring, multi-domain, privacy protecting and technology agnostic ML pipeline orchestration can be achieved by deploying a set of hierarchical MLFOs.

## 4.1 Problem overview

Consider a network operator providing 5G/B5G network services to a smart factory using a private 5G network. The RAN is shared between private and public users[20]. Public users may include general subscribers outside the smart factory, while private users are devices inside the smart factory which are typically autonomous robots, autonomous forklifts etc. These devices often have limited compute and run on battery power and hence are unable to run compute intensive ML tasks. Thus these ML tasks can be offloaded either to edge, central cloud or other devices inside factory with required compute and energy capabilities. In order to achieve dynamic offloading of these tasks based on network conditions, it is necessary to have a ML pipeline orchestrator which can orchestrate the required ML pipelines in autonomic fashion. This scenario also requires ML pipeline orchestration across multiple domains which are administered by different entities (i.e. mobile network operator, factory owner).

## 4.2 Hierarchical MLFOs in smart factory

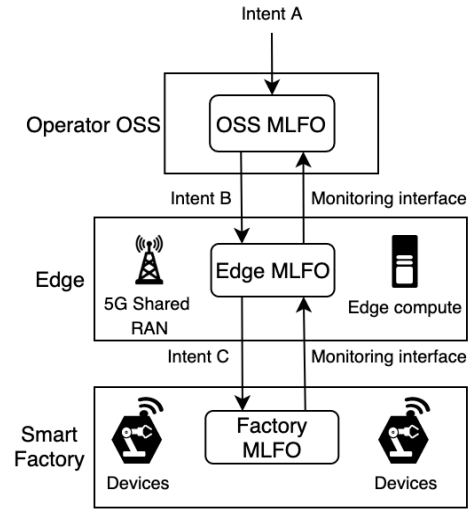

Fig. 3 – Smart Factory with hierarchical MLFO

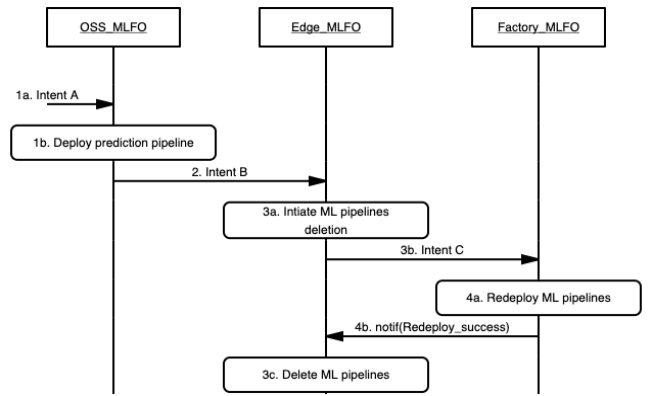

Fig. 4 – Workflow sequence diagram

We enable the dynamic offloading of ML tasks by deploying a set of hierarchical MLFOs in smart factory network. As seen in figure 3, we assign one MLFO each for OSS, edge and smart factory domains. These MLFOs communicate using two interfaces, an intent based interface and a monitoring interface. Figure 4 shows the sequence diagram for ML pipeline orchestration workflow. The ML pipeline orchestration workflow is as follows:

```
intentid: intent_a
targets:
    - id: OSS-01
      operation: maintain
      operand: SLA
      oparams:
        network: Edge-smart-factory-01
```
Listing 1: Intent A

1. The network operator provides intent A (operator intent) to the OSS MLFO. This intent instructs it to maintain the Service Level Agreements (SLA) for the smart factory network. In response to this the OSS MLFO deploys a ML pipeline to predict if there will be any QoS deterioration in the future.

```
intentid: intent_b
targets:
   - id: Edge-smart-factory-01
      operation: anticipate
      operand: QoS
      oparams:
         expected_drop: 20%
   - id: Edge-smart-factory-01
      operation: maintain
      operand: ml_inference
      oparams:
         minaccuracy: 95%
```
Listing 2: Intent B

2. The OSS using its ML pipeline is able to predict that QoS experienced by the private users in the smart factory edge is going to deteriorate due to increase in number of public users. Hence it sends intent B (MLFO intent) to the Edge MLFO instructing it to anticipate the drop in QoS and to maintain the accuracy of running ML inference tasks.

```
intentid: intent_c
targets:
   - id: Factory-01
      operation: stop
      operand: ml_offload
```
Listing 3: Intent C

3. Edge MLFO has some ML tasks which are offloaded from the devices inside the smart factory. On receiving intent B, it recognises that the sudden drop in the network QoS might reduce the performance of these tasks. It then prepares to tear down all the ML pipelines related to the offloaded tasks and sends intent C (MLFO intent) to factory MLFO. This intent instructs it to stop offloading the ML tasks. Once the factory MLFO orchestrates pipelines in its local domain for the relevant tasks, edge MLFO tears down all the offloaded ML pipelines.

4. Factory MLFO receives intent C, it identifies affected ML pipelines and redeploys them locally. It notifies edge MLFO about the redeployment. It may decide to run ML tasks locally on the devices or offload it to other devices inside the factory using networks independent from RAN (e.g WiFi, mm wave).

## 4.3   Intent conflict resolution

Conflicting intents requests from other MLFO entities might be handled by the policy management module of the MLFO. This might be done by using predefined policies (e.g. *request from MLFO X always get higher preference*) or using a more sophisticated conflict resolution algorithms. These algorithms may consider various parameters like priority of intent, hierarchy rank of MLFO (higher, lower) etc. Strategies used for such algorithms are currently not in scope of our work. It is worth noting that in some scenarios human-in-the-loop approach might be more suitable for conflict resolution.

## 4.4   Result

Our solution achieves the following qualitative results in above use case. We intend to publish the further quantitative results in a followup publication subject to completion of our testbed.

1. Self-configuring ML pipeline orchestration: ML pipelines in the network can be dynamically deployed, redeployed and teared down without human intervention depending on the network conditions and the requirements.

2. Multi-domain ML pipeline orchestration: ML pipelines can be orchestrated across multiple domains even though they are separately owned and administered by mobile operator and smart factory.

3. Privacy preservation: MLFOs in network operator's domain do not need to know network architecture or other internal complexities of smart factory because of the intent based interfaces.

4. Technology agnostic: ML pipeline orchestration can be performed irrespective of the underlying radio access technology or ML framework used in the smart factory.

Note that although we describe the results with respect to smart factory use case, similar results can be achieved when our solution is applied to 5G/B5G use cases in other verticals. Especially the ones with significant privacy concerns such as smart healthcare, smart construction, smart cities etc.

## 5.   CONCLUSION

We present a framework for autonomic orchestration of ML pipelines in mobile networks. Although our solution does not make ML pipeline orchestration fully autonomic, it takes a step in the right direction by enabling self-configuration. It also allows ML pipeline orchestration to be done in privacy-preserving and technology agnostic fashion. We provide a concrete example of our solution by applying it to a smart factory use case.

A major advantage of our framework is that it extends a globally standardised framework for ML pipeline orchestration. This makes our work more compelling rather than novel solutions developed in research silos.

## AUTHORS

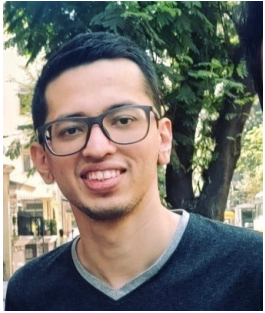

**Abhishek Dandekar** received his bachelor's degree in Telecommunication Engineering from University of Mumbai in 2015. He is currently pursuing his master's degree in ICT innovation at TU Berlin and writing his master's thesis at Fraunhofer HHI. Before this, Abhishek was with information networking lab at IIT Bombay, where he worked on developing 5G solutions for rural India using softwarised WLAN networks. He has contributed to IEEE and ITU standardisation working groups and holds a patent for controlling SDN based multi-RAT network. He won judges' prize in ITU AI/ML in 5G 2020 challenge. His current research interests include autonomous networks, distributed ML orchestration and industrial 5G.